\documentclass[aps,pre,twocolumn,showpacs,superscriptaddress]{revtex4}

\usepackage[latin1]{inputenc}
\usepackage[english]{babel}
\usepackage{bm,graphicx,graphics,amsmath,amssymb,epsfig,color}

\usepackage[colorlinks,citecolor=blue,linkcolor=cyan]{hyperref}
\usepackage[usenames,dvipsnames,svgnames,table]{xcolor}

\def \be {\begin{equation}}
\def \ee {\end{equation}}
\def \beA {\begin{eqnarray}}
\def \eeA {\end{eqnarray}}

\def \der {\partial}

\def \Re  {\rm{Re}}
\def \Im  {\rm{Im}}
\def \average#1{\left\langle #1 \right\rangle}

\def \graffb#1{\left\{ #1 \right\}}

\begin{document}
\title{Stochastic thermodynamics of oscillators networks}
\author{Simone Borlenghi} 
\affiliation{Department of Applied Physics, School of Engineering Science, KTH Royal Institute of Technology, Electrum 229, SE16440 Kista, Sweden}
\author{Anna Delin}
\affiliation{Department of Applied Physics, School of Engineering Science, KTH Royal Institute of Technology, Electrum 229, SE16440 Kista, Sweden}
\affiliation{Swedish e-Science Research Center (SeRC), KTH Royal Institute of Technology, SE-10044 Stockholm, Sweden}
\begin{abstract}
We apply the stochastic thermodynamics formalism to describe the dynamics of systems of complex Langevin and Fokker-Planck equations. We provide in particular a simple and general recipe to calculate
thermodynamical currents, dissipated and propagating heat for networks of nonlinear oscillators. By using the Hodge decomposition of thermodynamical forces and fluxes, we derive a formula for entropy production that generalises the notion of non-potential forces and makes transparent the breaking of detailed balance and of time reversal symmetry for states arbitrarily far from equilibrium. Our formalism is then applied to describe the off-equilibrium thermodynamics of a few
examples, notably a continuum ferromagnet, a network of classical spin-oscillators and the Frenkel-Kontorova model of nano friction.
\end{abstract}
\maketitle 

\section{introduction}
Dissipation and heat transfer are universal phenomena in Physics, appearing whenever a small system is coupled to the much larger environment. In this situation, it is in practice not possible keep track of all the observables of the universe,
instead some measurable macroscopic quantities, such as energy, entropy and heat flows are used to describe the evolution of the system and of its average properties.

In the presence of several environments (or thermal baths/reservoirs) at different temperatures, the system reaches a non-equilibrium steady state where thermodynamical currents (such as heat, energy, spin, electrical)
may flow through the system from a reservoir to the other. Close to equilibrium, those currents are proportional to the corresponding thermodynamical forces. Examples of thermodynamical forces include gradients or differences of temperature, voltage, chemical potentials and concentrations of chemical species etc. The language of thermodynamical forces and currents, which is nowadays the cornerstone of non equilibrium thermodynamics, was first developed by L. Onsager in the 1930s \cite{onsager31a,onsager31b}, and by R.Kubo in the 1950s
\cite{kubo57}. The formalism can be naturally extended beyond the linear regime, as it was first observed by Schnakenberg \cite{schnakenberg76} and subsequently in more recent works on stochastic thermodynamics \cite{seifert12}.

In out-of-equilibrium setups, it is of primary importance to determine the (possibly many) currents that flow between subparts of the system, together with the corresponding forces, and the heat flow that is dissipated to the environment \cite{borlenghi17a}. The first case corresponds typically to the work done on the system by the environment, while the latter is associated to the production of entropy and an increased disorder of the ensemble (system+environment), and it is related to the efficiency of the thermodynamical process. In networks of nonlinear oscillators, the heat flow throughout the system is a coherent phenomenon that requires synchronisation of the oscillators, while the dissipated heat is incoherent \cite{borlenghi14a,borlenghi14b,borlenghi15,borlenghi17a}.

The scope of this paper is to provide a general recipe to calculate currents, dissipated heat and work in a large class of out-of-equilibrium systems. To this end, we shall adopt the formalism of stochastic thermodynamics (ST) \cite{seifert12} applied to the dynamics of complex-valued Langevin and Fokker-Panck equations.  At variance with the standard formulation of stochastic thermodynamics, which uses colloidal particles as paradigm \cite{seifert12}, here we provide a systematic way to describe the non-equilibrium thermodynamics of networks of nonlinear oscillators. Following the idea of ST, we start from the stochastic trajectories of a small ensemble of oscillators coupled to a bath and we extract useful information (such as currents and entropy) from the ensemble averages of the main observables. As the system evolves, different parts of the network and the environment become statistically correlated. 
The currents are expressed in terms of those correlations.

The present paper is organised as follows: in Sec.II we apply the Lagrangian and Hamiltonian formalism to describe complex-valued equations of motion, following 
Refs.\cite{dekker75,dekker77,dekker79,tekkoyun03}. In particular, we formulate the conservative and dissipative dynamics respectively in terms of Poisson commutators and anti-commutators for a possibly non-Hermitian Hamiltonian, a topic that has been extensively studied \cite{dekker75,morrison84,morrison86,rotter09,ottinger11}. 

In Sec.III we develop the stochastic thermodynamics formalism for complex-valued equations, and derive a simple and general formula for entropy production, which makes transparent the breaking of detailed balance and is proportional to the heat dissipated to the environment. This section generalises previous work \cite{borlenghi17a,tome06,tome10} to complex Langevin equations with multiplicative noise.

Sec. IV contains the formulation of the first principle of thermodynamics. This constitutes the main result of this paper and allows one to identify the heat transported and dissipated.

Sec. V provides some examples of realistic physical systems where thermodynamical currents and entropy production are calculated. We shall describe in particular the dynamics of a one dimensional continuum ferromagnet, of a network of classical magnetic spins and of the Frenkel-Kontorova model \cite{braun98} for nano-friction. Finally, the main results and conclusions are summarised in Sec. VI.

\section{Hamiltonian-Lagrange formulation for complex Langevin equations}%

We consider here the following complex Langevin equation:
\be\label{eq:langevin1}
\dot{\psi}_m=F_m+g_m\xi_m(t),
\ee
where the dot indicates time derivative and $\psi_m=\sqrt{p_m(t)}e^{i\phi_m(t)}$ is a complex
wave function with amplitude $p_m=|\psi_m^2|$ (also referred as power) and phase $\phi_n$. The force $F_m$ is an arbitrary function of the 
$\bm{\psi}=(\psi_1,...,\psi_N)$ and their complex conjugate $\bm{\psi}^*=(\psi_1^*,...,\psi_M^*)$.
We assume that both the coupling between the $\bm{\psi}$s and the damping are contained in the definition of $F_m$. 

The terms $\xi_m$, which model the stochastic baths, are complex Gaussian random processes with zero average and correlation  $\average{\xi_m(t)\xi^*_n(t^\prime)}=\delta_{mn}\delta(t-t^\prime)$.
Here $g_m$ is an arbitrary function of the $(\bm{\psi},\bm{\psi^*})$. Throughout the paper vectors and matrices are written in
bold text, while their components are written in plain text and denoted by the $m$ and $n$ subscripts. 
The quantity $|g_m^2|$ plays the role of diffusion constant. We assume that the latter is proportional a damping coefficient $\Gamma_m$ and temperature $T_m$, according to the fluctuation-dissipation theorem.
Thus, at variance with previous studies \cite{borlenghi17a}, we consider here the more general situation of Langevin equations with multiplicative noise.

The force is given by the derivative 
\be\label{eq:derivative}
F_m=i\der^*_m\mathcal{H}
\ee 
of a complex (and possibly non-Hermitian) Hamiltonian $\mathcal{H}$. Here the Wirtinger derivatives are defined as
\be\label{eq:wirtinger}
\der_m\equiv\frac{\der}{\der\psi_m}=\frac{1}{2}\left(\frac{\der}{\der x_m}-i\frac{\der}{\der y_m}\right)
\ee
where $\der_m^*$ is the complex conjugate and $\psi_m=x_m+iy_m$. The complex conjugate equation to Eq.(\ref{eq:langevin1}) contains the forces $F_m^*=-i\der_m\mathcal{H}$. 
A straightforward calculation shows that in a dissipative system with Hamiltonian $\mathcal{H}=\mathcal{H}^R+\mathcal{H}^I$ , where $R$ and $I$ are respectively the Hermitian (or reversible) and anti-Hermitian (or irreversible) components,
the dynamics of an arbitrary function $f$ of the observables $(\bm{\psi},\bm{\psi}^*)$ can be written as

\be\label{eq:commutators1}
\dot{f}=i\graffb{f,\mathcal{H}^R}_-+\graffb{f,\mathcal{H}^I}_+.
\ee
Here the Poisson commutators ($-$) and anti-commutators($+$) are defined respectively as
\be\label{eq:commutators2}
\graffb{f,\cdot}_\mp=\sum_m \left(\frac{\der f}{\der \psi_m^*}\frac{\der}{\der\psi_m}\mp\frac{\der f}{\der\psi_m}\frac{\der}{\der\psi_m^*}\right).
\ee
We note in particular that, from Eqs.(\ref{eq:derivative}-\ref{eq:commutators1}), the reversible and irreversible forces can be expressed as $F_m^R=i\{\mathcal{H}^R,\psi_m\}_-$ 
and $F_m^I=\{\mathcal{H}^I,\psi_m\}_+$. On the other hand, the couple $(\psi_n,i\psi^*_n)$
are canonical conjugate variables, since one has $i\{\psi_m,i\psi_n^*\}_-=\delta_{mn}$.

Since the commutators and anti-commutators define respectively a symplectic and a metric structure on the space tangent to the phase space, this kind of system is called metriplectic.
The formulation of dissipative dynamics in terms of anti-brackets in metriplectic structures has been extensively studied, both for classical and quantum systems \cite{morrison86,ottinger97a,ottinger97b,guha07}. In those formulations, the irreversible part of the Hamiltonian is usually identified with the entropy of the system. Here we do not pursue this identification, since we will describe the irreversibility in terms of the information entropy and the associated entropy production, as it is customary in the
ST formalism. Later in the paper, we shall elucidate the connection between the irreversible part of the Hamiltonian and the entropy production.

An important step here is to determine the canonical transformations, that must preserve the metriplectic structure.  
In practice, from the definition of force and from Eqs.(\ref{eq:commutators2}) one must have that
\be\label{eq:canonical1}
\dot{b}_k=i\graffb{\mathcal{H}^R,b_k}_-+\graffb{\mathcal{H}^I,b_k}_+=\frac{\der}{\der b^*_k}(\mathcal{H}^R+\mathcal{H}^I)
\ee
for a variable $b_k$ function of the old coordinates $\bm{\psi}$. From the chain rule of partial derivative one has

\be\label{eq:canonical2}
\frac{\der}{\der b^*_k}= \sum_m\left(\frac{\der \psi_m^*}{\der b_k^*}\frac{\der }{\der{\psi^*_m}}+\frac{\der \psi_m}{\der b_k^*}\frac{\der }{\der{\psi_m}}\right)
\ee
However, from the definition of commutators and anti-commutators the following equalities must also hold:
\beA\label{eq:canonical3}
\frac{\der}{\der b^*_k}=\sum_m\left(\frac{\der b_k}{\der \psi_m}\frac{\der }{\der{\psi_m^*}}-\frac{\der b_k}{\der \psi^*_m}\frac{\der }{\der{\psi_m}}\right)\\
\frac{\der}{\der b^*_k}=\sum_m\left(\frac{\der b_k}{\der \psi_m}\frac{\der }{\der{\psi_m^*}}+\frac{\der b_k}{\der \psi^*_m}\frac{\der }{\der{\psi_m}}\right)
\eeA
The two equalities can be both satisfied only if the following holds:

\beA\label{eq:canonical4}
\frac{\der b_k}{\der \psi_m}  &=&\frac{\der \psi_m^*}{\der b_k^*}\\
\frac{\der b_k}{\der \psi^*_m} &=&\frac{\der \psi^*_m}{\der b_k}=0.
\eeA
This means essentially that the new coordinates must be analytic functions of the old ones, since they cannot contain both a variable and its complex conjugate. 
Adding a complex number or performing a $U(1)$ gauge transformation preserves the commutators \cite{borlenghi16a}, however note that the Bogoliubov transformations \emph{are not} canonical in this case, although they are canonical transformations of the system without dissipation.

Note that the system can be described using the following Lagrangian, similar to the one for the heat equation \cite{dekker77}:

\be\label{eq:lagrangian}
\mathcal{L}=\frac{i}{2}\sum_m\left(\dot{\psi}_m\psi_m^*-\dot{\psi}^*_m\psi_m\right)-\mathcal{H}.
\ee
The equations of motion for $\psi^*_m$ are given by the Euler-Lagrange equations

\be\label{eq:euler_lagrange}
\frac{d}{dt}\frac{\der\mathcal{L}}{\der\dot{\psi}_m}-\frac{\der \mathcal{L}}{\der \psi_m}=0
\ee
while the dynamics of $\bm{\psi}$ given by the complex conjugate equations. Eqs.(\ref{eq:lagrangian}) and (\ref{eq:euler_lagrange}) are particularly useful to determine the conserved currents of the system associated to the invariance of the Lagrangian with respect to a global $U(1)$ transformation, as it will be clarified in the next sections.

\section{Fokker-Planck equation and entropy production}%
This section generalises the material presented in Ref. \cite{borlenghi17a} to the case of multiplicative noise.
The time evolution of the probability distribution oin the phase space, associated to the Langevin Eq.(\ref{eq:langevin1}), is given by the following
Fokker-Planck (FP) equation: 
\be\label{eq:fp1}
\dot{P}=\sum_m\left[ -\der_m(F_mP)-\der_m^*(F_m^*P)+2\der_m\der_m^*(|g_m|^2P)\right].
\ee
Following Refs.\cite{tome06,borlenghi17a}, we define the reversible and irreversible probability currents as
\beA\label{eq:probcurr}
\mathcal{J}^I_m &=& F_m^IP-D_m\der_m^*(|g_m|^2P)\nonumber\\
\mathcal{J}^R_m &=& F_m^RP,
\eeA
with $\mathcal{J}_m=\mathcal{J}_m^R+\mathcal{J}_m^I$ and $\mathcal{J}_m^*$ the complex conjugate.
By using those currents, Eq. (\ref{eq:fp1}) assumes the usual form of a continuity equation:

\be\label{eq:fpcont}
\dot{P}=-\sum_m[\der_m\mathcal{J}_m-\der_m^*\mathcal{J}_m^*].
\ee
Thermal equilibrium corresponds to the case where the probability currents are zero, while non-equilibrium steady state corresponds to non-zero divergenceless currents, whith $\dot{P}$=0.

The entropy flow $\Phi$ and entropy production $\Pi$ are obtained starting from the definition of phase space entropy 
\be\label{eq:entropy0}
S=-\average{\log P}\equiv -\int P \log P dx,
\ee
where $\average{\cdot}$ denotes the ensemble average.
Computing the time derivative $\dot{S}$ by means of Eq.(\ref{eq:fpcont}) we obtain:

\be\label{eq:entropy1}
\dot{S} =\int\sum_m(\der_m\mathcal{J}_m+\der^*_m\mathcal{J}^*_m)\ln Pdx.
\ee 
Upon integrating by parts, and assuming that the reversible forces have zero divergence, Eq.(\ref{eq:entropy1}) 
becomes 
\be\label{eq:entropy2}
\dot{S}=-2{\Re}\int\sum_m \mathcal{J}^I_m\frac{\partial_m P}{P}dx.
\ee
From the definition of probability currents Eq.(\ref{eq:probcurr}) one has

\be\label{eq:logp}
\frac{\der^*_mP}{P} = \frac{-\mathcal{J}^I_m}{P|g_m|^2}-\frac{F_m^I}{|g_m|^2}-\der_m^*\ln |g_m|^2,
\ee
together with the complex conjugate equation. Upon substituting the previous equation into Eq.(\ref{eq:entropy2}) gives

\beA\label{eq:entropyprod1}
\dot{S} &=&2{\Re}\int\frac{\sum_m\mathcal{J}_mF_m^{I*}}{|g_m|^2}dx-2\int \sum_m\frac{|\mathcal{J}^I|^2}{P|g_m|^2}dx\nonumber\\
	    &-&2{\Re}\int\sum_m\mathcal{J}_m^I\der_m\ln|g_m|^2dx.
\eeA
The first and second terms are respectively entropy flow and entropy production. We remark that we here consider only steady states. In this condition, assuming that the probability currents vanish at infinity \cite{tome06,borlenghi17a}, one can integrate by part the last term, which is proportional to $\der_m\mathcal{J}+c.c.$. However, since the divergence of the thermodynamical currents is zero in stationary states, the last term vanishes in that case. Thus, in steady state the entropy flow $\Phi$ is minus the entropy production $\Pi$ \cite{tome06,borlenghi17a}, as in the case of additive noise.

At this point we substitute integrals containing $P$ with ensemble averages. In this way Eqs.(\ref{eq:probcurr}) and (\ref{eq:entropyprod1}) give
\be\label{eq:entropyprod2}
\Phi =2\sum_m\average{\frac{|F_m|^2}{|g_m|^2}}+2{\Re}\sum_m\average{\der_mF_m^I},
\ee
which is the same expression obtained in Ref.\cite{borlenghi17a}, with $|g_m|^2$ playing the role of diffusion constant. As in Refs.\cite{tome06,borlenghi17a} we identify the quantity $T\Phi$ with the heat exchanged with the bath.

We proceed now by deriving an expression for the entropy production that makes transparent the breaking of detailed balance and the onset of irreversibility.
Since in steady states one has $\sum_m(\der_m\mathcal{J}^*_m+\der_m\mathcal{J^*}_m)=0$, one can apply the Hodge decomposition \cite{wells80} and write the currents as
\be\label{eq:hodge}
\mathcal{J}^*_m=\sum_{\ell}\der_\ell\Omega_{\ell m}+\der_m\Lambda, 
\ee
where $\Omega$ is an anti-symmetric tensor and $\Lambda$ a scalar. 
We separate the entropy flow into two components $\Phi_1$ and $\Phi_2$ containing respectively $\Omega$ and $\Lambda$. For the first component one has
\beA\label{eq:forces}
\Phi_1 &=&2{\Re}\int\sum_{\ell}\der_\ell\Omega_{\ell m}\frac{F_m}{|g_m|^2}dx\nonumber\\
       &=&2{\Re}\int\sum_{ell}\Omega_{\ell m}\left(\frac{\der_\ell F_m}{|g_m|^2}-\frac{\der_m F_\ell}{|g_\ell|^2}\right)dx
\eeA
where we have used the anti-symmetry of $\Omega$ and integrated by parts discarding the boundary terms.
One has that the condition of detailed balance is $\frac{\der_m F_\ell}{|g_\ell|^2}-\frac{\der_\ell F_m}{|g_m|^2}=0$, which is met when the forces are potentials and/or the 
temperatures are the same, $|g_m|^2=|g_\ell|^2$. Note that this condition generalises the formulation of Refs.\cite{tome06,tome10} to the case of complex-valued forces.

However, in our system we have two coupled currents, associated respectively to the conservation of energy and of the total power $p_n$, or "number of particles". 
Thus, we expect that the entropy production contains two components: one that depends on the temperature differences and one that depends on the chemical potential differences.

To see this, let us write the force as the derivative of the following Hamiltonian:
\be\label{eq:ham}
\tilde{\mathcal{H}}=\mathcal{H}+i\sum_k\mu_k|\psi_k|^2 
\ee
where $\mu_k$ is the local chemical potential.

A straightforward calculation gives for Eq.(\ref{eq:forces}):
\beA
\Phi_1&=&2{\Re}\int \sum_{\ell m}\Omega_{\ell m}\left(\frac{\der_\ell\der_m^*}{|g_m|^2}-\frac{\der_m\der_{\ell}^*}{|g_\ell|^2}\right)\mathcal{H}dx\nonumber\\
           &+&2{\Re}\int \sum_{\ell m}\Omega_{\ell m}\left(\frac{\mu_m}{|g_m|^2}-\frac{\mu_\ell}{|g_\ell|^2}\right)dx.
\eeA
The first term is non zero if the Hamiltonian is non-Hermitian and/or if the temperatures are different. On the other hand, the second term is non zero if the chemical potentials or the temperatures are different.

There is also another way to drive the system off equilibrium: by applying a \emph{constant} chemical potential that compensates the damping \cite{slavin09,borlenghi17a}. In this case one expects that the entropy production is non zero, even if the current vanishes.
To see this, we consider the second contribution to the entropy production:
\beA\label{eq:phi2_ham}
\Phi_2 &=& -2{\Re}\int\der_m\Lambda F_m dx=\nonumber\\
           &=& 2{\Re}\int i \Lambda\der_m\der_m^*[\mathcal{H}-\sum_k\mu_k |\psi_k|^2]dx\nonumber\\
           &=& 2{\Re}\int i \Lambda(\der_m\der_m^*\mathcal{H}-\mu_m)dx\nonumber\\
\eeA
However, if we write the Hamiltonian as in the case of the DNLS \cite{iubini13,borlenghi17a}, the term ${\Re}[\der_m\der_m^*\mathcal{H}]$ is the damping of the system, $\Gamma_m$. Thus one has $\Phi_2\propto \int i\Lambda (\Gamma_m-\mu_m)dx$.
This shows that the system does not relax to equilibrium in the case where the chemical potential compensates the damping, as has been pointed out also in Ref.\cite{slavin09}.

\section{Transported vs dissipated heat}%

This section contains the main results of the paper. Starting from the first principle of thermodynamics, we derive the expressions for the heat dissipated and flowing through the system.
 To keep the notation simple, we consider the case with $g_m=D_m\equiv\alpha T_m$, with $\alpha$ the damping of the system. It is straightforward to generalise our discussion to the case of multiplicative noise.
Following Ref.\cite{borlenghi17a}, for a network of $m=1,...,M$ oscillators, we consider the Hamiltonian 
\be
\mathcal{H}=\sum_m[h_m+i\alpha(h_m+\mu|\psi_m|^2)]
\ee
where $h_m$ is the local energy, and the Hamiltonian splits into a reversible and irreversible component, respectively $\mathcal{H}^R=\sum_mh_m$ and $\mathcal{H}^I=i\alpha\sum_m(h_m+\mu_m|\psi_m|^2)$. Here $\mu_m$ is the local chemical potential, while
$|\psi_n|^2$ plays as usual \cite{borlenghi17a} the role of particle number.
 
The first principe of thermodynamics can be expressed as a balance equation for the energy according to

\be\label{eq:firstprinciple}
\frac{d}{dt} \frac{1}{i\alpha}\average{\mathcal{H}^I}=\frac{1}{i\alpha}\int \dot{P}\mathcal{H}^Idx+\frac{1}{i\alpha}\int P\dot{\mathcal{H}}^Idx
\ee

where $dx=\frac{i}{2}\sum d\psi_m\wedge d\psi_m^*$ is the volume element of the phase space. In Ref.\cite{seifert12}  and in stochastic thermodynamics in general, the first and second terms of the previous equation are respectively
identified with heat $Q$ and work $W$. However, in the present case one does not have a clear distinction between heat and work. Instead, one can differentiate the heat propagating through the system from the heat
from the heat dissipated to the bath. To see this, we started by calculating the value of $Q$.
We remark that, as observed in Ref.\cite{borlenghi17a}, only the irreversible part of the Hamiltonian enters these expressions.
We use the FP equation Eqs.(\ref{eq:fp1}) and (\ref{eq:fpcont}) and substitute $\dot{P}$ with the derivative of the currents $\mathcal{J}$: 
\be\label{eq:heat2}
Q = \frac{1}{i\alpha}\int\sum_m (\der_m\mathcal{J}+\der_m^*\mathcal{J}^*)\mathcal{H}^Idx
\ee
Then, upon substituting the expression for the currents, integrating by parts and discarding boundary terms one has
\beA\label{eq:Q}
Q&=&-\frac{2}{i\alpha}{\Re}\int\sum_m(F_m^IP-D_m\der_m^*P)\der{\mathcal{H}^I}dx\nonumber\\
&=&\sum_m\left( \frac{2}{\alpha} \average{|F_m^I|^2}+2\frac{D_m}{\alpha}{\Re}\average{\der_mF_m^I}\right).
\eeA

This corresponds to the entropy flow multiplied by the temperature, which constitutes the heat dissipated to the environment.This generalises to the complex case the results obtained
in Refs.\cite{tome06,tome10}.

From Eq.(\ref{eq:firstprinciple}), we calculate $W$ as
\be\label{eq:heat5}
W =\frac{1}{i\alpha}\int P \dot{\mathcal{H}}^Idx=\frac{1}{i\alpha}\sum_m\average{\der_m\mathcal{H}^I\dot{\psi}_m+\der_m^*\mathcal{H}^I\dot{\psi}^*_m}.
\ee
Applying the substitution $\der_m\mathcal{H}^I=iF^*_m$ and its complex conjugate and substituting $\dot{\psi}_m$ with the equation of motion, a straightforward calculation shows that
$W=j_m^Q+Q$, where
\be\label{eq:heatflow}
j_m^Q=\frac{2}{\alpha}{\Re}\average{F_m^RF_m^{*I}}. 
\ee
is the heat flowing through the mth oscillator, and therefore transported along the chain \cite{borlenghi17a}.

\section{Application to physical systems}%

\subsection{Hamilton-Lagrange description of a one dimensional continuum ferromagnet} %

\begin{figure}
\begin{center}
\includegraphics[width=8.0cm]{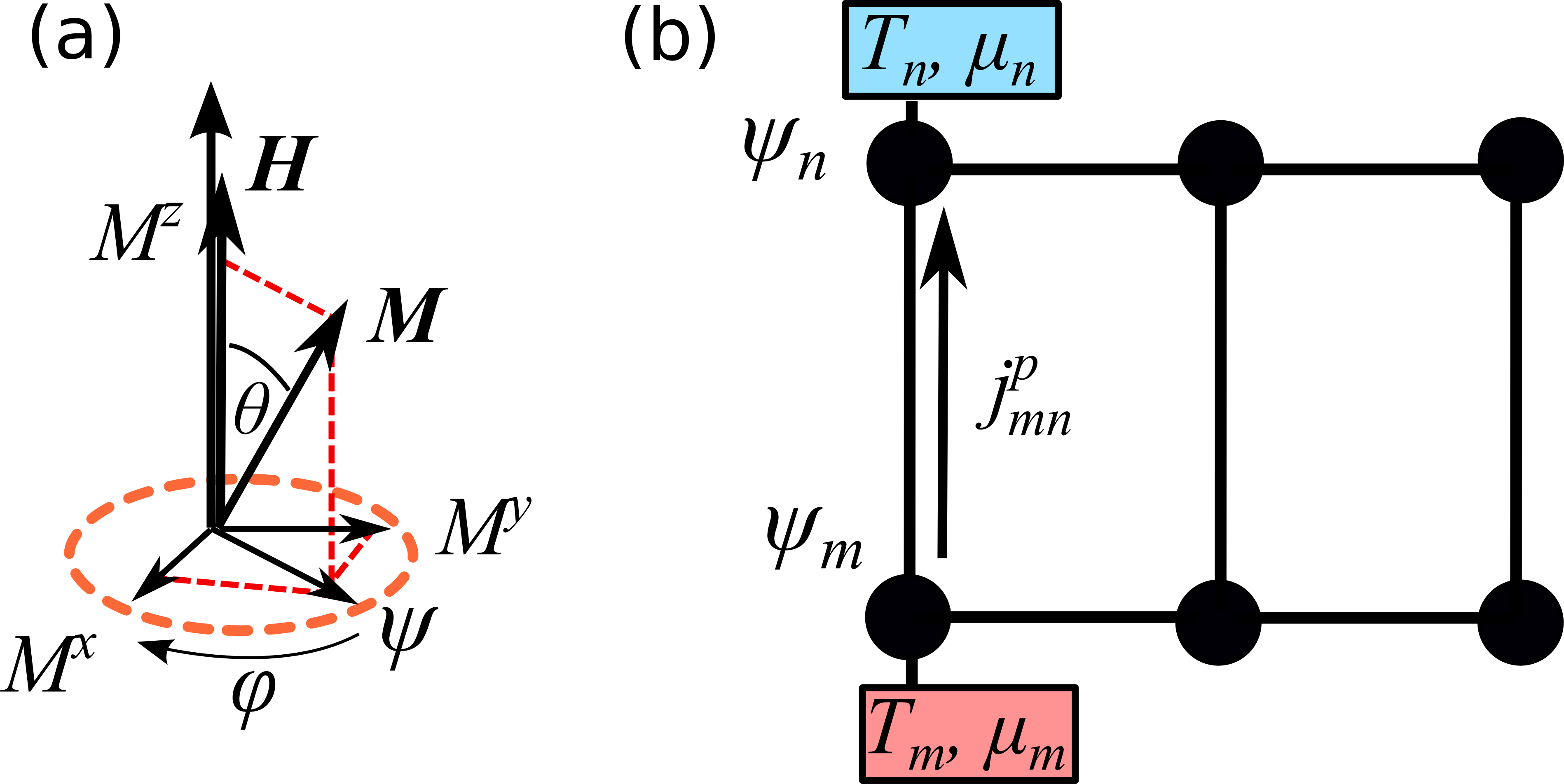}
\caption{ a) Magnetisation vector $\bm{M}$ precessing around the effective field $\bm{H}$ along the $z$ direction. The precession occurs in the $x$-$y$ plane and is conveniently described by the stereographic projection $\psi$.
b) Network of nonlinear oscillators connected to thermochemical baths with different temperatures and chemical potentials. The "particle" current $j_{mn}^p$ describe the transport of the local power $p_m$ between oscillators $m$ and $n$.}
\label{fig:figure1}
\end{center}
\end{figure}

We consider here the dissipative dynamics of a continuum ferromagnet at zero temperature. In particular, we show how the symmetry of the system allows one to obtain the conserved quantities and the corresponding equations of motion.

The (linearised) magnetisation dynamics close to equilibrium of such a system is described by the following Schr\"odinger equation with complex potential \cite{lakshmanan11}
\be\label{eq:schroedinger}
\dot{\psi}=-(\Gamma-\mu)\psi-i\omega\psi-iA\der_x^2\psi.
\ee
The stereographic variable $\psi(x,t)=(M_x+iM_y)/2M_s$ describes the precession of the magnetisation vector $\bm{M}=(M_x,M_y,M_z)$ in the $x$-$y$ plane, around the $z$ axis, see Fig.\ref{fig:figure1}a).  Here $M_s$ is the saturation magnetisation. 
The precession frequency reads $\omega=\gamma h_{\rm{ext}}$,
where $\gamma$ is the gyromagnetic ratio and $h_{\rm{ext}}$ is the applied field along the $z$ direction. The quantity $\Gamma=\alpha\omega$ is the damping rate, proportional to the phenomenological damping parameter $\alpha$.
The chemical potential $\mu$ accounts for spin transfer torque, which can compensate the damping and leads to a steady state precession of the magnetisation \cite{slavin09}. The spin stiffness $A$ is the strength of the exchange interaction.

Note that in realistic cases one should consider a nonlinear damping \cite{slavin09,iubini13} $\Gamma(p)\approx\Gamma_0(1+2p)$, with $p\equiv |\psi|^2$, that allows the system to have limit cycle oscillations when $\mu>\Gamma_0$. 
Apart from chemical potential term, Eq.(\ref{eq:schroedinger}) can have more terms, accounting for temperature and additional time-dependent magnetic  fields that drive the system out of equilibrium.
In the present case however we shall first consider the linearised dynamics with zero temperature, since this is sufficient for our purpose to derive and illustrate the expressions for the spin currents. The more general case of a network at finite temperature will be discussed in the next section.

The Lagrangian density for our Schr\"odinger equation reads

\be\label{eq:lagrangian}
\mathcal{L}=\frac{i}{2}(\dot{\psi}\psi^*-\psi\dot{\psi}^*)-(i\Gamma+\omega)|\psi|^2-A\der_x\psi\der_x\psi^*.
\ee
The equations of motion are given by the Euler-Lagrange equations

\be\label{eq:eulag}
\frac{\delta\mathcal{L}}{\delta\psi^*}\equiv \frac{d}{dt}\frac{\der\mathcal{L}}{\der\dot{\psi}^*}+\der_x\frac{\der\mathcal{L}}{\der(\der_x\psi^*)}-\frac{\der \mathcal{L}}{\der\psi^*}=0
\ee
with the dynamics for $\psi^*$ being given by $\frac{\delta \mathcal{L}}{\delta\psi}=0$.
From the Lagrangian one can derive the following moments

\beA\label{eq:momenta}
\Pi_t &=& \frac{\der\mathcal{L}}{\der\dot{\psi}}=\frac{i}{2}\psi^*\nonumber\\
\bar{\Pi}_t &=& \frac{\der\mathcal{L}}{\der\dot{\psi}^*}=-\frac{i}{2}\psi\nonumber\\
\Pi_x &=& \frac{\der\mathcal{L}}{\der(\der_x\psi)}=-J\der_x\psi^*\nonumber\\
\bar{\Pi}_x &=& \frac{\der\mathcal{L}}{\der(\der_x\psi^*)}=-J\der_x\psi
\eeA
and finally a Legendre transform gives the following complex Hamiltonian density:

\beA\label{eq:hamiltonian}
\mathcal{H} &=& \Pi_t\dot{\psi}+\bar{\Pi}_t\dot{\psi}^*-\mathcal{L}\nonumber\\
                    &=& (\omega-i\Gamma)|\psi|^2+A|\der_x\psi|^2.
\eeA
From the Hamiltonian one obtains the equation of motion Eq.(\ref{eq:schroedinger}) as $\dot{\psi}=\frac{\delta\mathcal{H}}{\delta i\psi^*}$, so that $\psi$ and $i\psi^*$ are conjugate variables. One can check that this 
equivalent to the derivation of the equations of motion using commutators and anti-commutators as described in Eq.(\ref{eq:commutators1}).

The conservation equation for the local spin wave power $p\equiv |\psi|^2$ is obtained from the invariance of the Lagrangian Eq.(\ref{eq:lagrangian}) with respect to the global phase transformation $\psi\rightarrow e^{-i\alpha}\psi$, with the
corresponding infinitesimal transformation $\delta\psi\approx-i\alpha\psi$. The invariance of the Lagrangian with respect to such infinitesimal transformation yelds $\frac{\delta\mathcal{L}}{\delta\psi}\delta\psi+c.c.=0$, $c.c.$ indicating the complex conjugate.
A straightforward calculation gives then
\beA\label{eq:conservation1}
0 &=& \frac{\delta\mathcal{L}}{\delta\psi}\delta\psi +c.c.\nonumber\\
  &=& \frac{\der\mathcal{L}}{\der\psi}\psi+\frac{\der\mathcal{L}}{\der\dot{\psi}}\dot{\psi}+\frac{\der\mathcal{L}}{\der(\der\psi)}\der\psi+c.c.
\eeA
By using Eqs.(\ref{eq:conservation1}), (\ref{eq:schroedinger}) and (\ref{eq:lagrangian}) one obtains the following conservation equation for the spin wave power
\be\label{eq:conservation2}
\dot{p}=-2(\Gamma-\mu)-\der_x j^p,
\ee
where the spin current reads $j^p=2A{\Im}[\psi^*\der_x\psi]$, while $\Gamma$ and $\mu$ act respectively as sink and source of excitations. We remark that this is precisely the same expression as the probability currents that appears in the Schr\"odinger equation of quantum mechanics. In the present case, it describes the transport of the $z$ component of the magnetisation along the system. Indeed, one can check that $j^p$ is the same as the spin-wave current 
$\bm{j}=A\bm{M} \times \nabla \bm{M}$ written in terms of the stereographic variable $\psi$ \cite{kajiwara10,borlenghi15a}





\subsection{Entropy production for a network of classical spins}%
The finite-temperature dynamics of an ensemble of magnetic spins $\{\bm{M}_m\}$, $n=1,...,M$, inside a ferromagnet is described by the Landau-Lifshitz-Gilbert (LLG) equation of motion \cite{gurevich96} with stochastic thermal baths. 
The LLG equation is a vector equation, and obtaining the associated FP equation in practice very cumbersome.
A great simplification is obtained by re-writing the LLG equation In terms of the complex variable $\psi_m=\frac{m_{xm}+im_{ym}}{1+m_{zm}}$, where $\bm{m}=\bm{M}/M_s$ is the magnetisation vector normalised over the saturation magnetisation.
In this way one obtains \cite{slavin09,lakshmanan11,borlenghi15a}

\be\label{eq:llgstereo}
\dot{\psi}_m=\frac{i+\alpha}{1+\alpha^2}\left(F_m+\sum_{k=1}^3g^{k}_m\xi^k_m\right),
\ee
where the force reads
\be\label{eq:llgforce}
F_m=\gamma H_z\psi_m+C\frac{1-|\psi_m|^2}{1+|\psi_m|^2}\psi_m+\sum_\ell A_{m\ell}\psi_\ell
\ee
The first term corresponds to the applied field $H_z$ along the precession axis $z$ of the magnetisation. The second term corresponds to the demagnetising field, while the last term models the coupling with the other spins.
Note that the formulation of the coupling is completely general. In particular, such coupling can have different origins (exchange or dipolar interaction) depending on the coupling matrix $A$, which can be a function of the $\bm{\psi}$s.
The reversible and irreversible components of the forces read respectively $F^R=\frac{i}{1+\alpha^2}F$ and $F^I=\frac{\alpha}{1+\alpha^2}F$.

The term $g^k_n$ in Eq.(\ref{eq:llgstereo}) is the strength of the noise, and models thermal fluctuations on site $n$. There are three components of the noise on each site, one per each direction of the magnetisation:
\beA\label{eq:baths}
g^1_n &=&\frac{1}{2} \sqrt{D_mT_m}(1-\psi_m^2),\nonumber\\ 
g^2_n &=&-\frac{i}{2}\sqrt{D_mT_m}(1+\psi_m^2),\nonumber\\
g^3_n &=&\sqrt{D_mT_m}\psi_m.
\eeA
Here $D_m=\frac{\alpha k_B}{\mu_0V_mM_s}$ is the diffusion constant, with $kB$ the Boltzmann constant, $\mu_0$ the vacuum magnetic permeability and $V_m$ the elementary volume containing the magnetisation vector at site $m$, of the order of few nm$^3$. $T_m$ is the temperature at site $m$. The $\bm{\xi}$ are Gaussian random variables with zero average and correlation $\average{\xi_m^k(t)\xi_{m\prime}^{k\prime}(t\prime)}=\delta_{{kk^\prime}{mm^\prime}}\delta(t-t^\prime)$

The entropy production splits into the sum of two components, $\Phi=\Phi_1+\Phi_2$, with
\be\label{eq:phi1}
\Phi=\frac{2\alpha^2}{1+\alpha^2}\average{\frac{\sum_m^2|F_m|^2}{\sum_{km}|g_m^k|^2}}+\frac{2\alpha}{1+\alpha^2}{\Re}\sum_m\average{\der_m F_m}
\ee
and 
\be\label{eq:phi2}
\Phi_2=\frac{2\alpha}{1+\alpha^2}\gamma h_z.
\ee

In the more general case where easy axis anisotropy and demagnetising field along $z$ are present, the LLG equation is still given by Eq.(\ref{eq:llgstereo}), but with reversible forces \cite{lakshmanan11}
\beA\label{eq:FR} 
F^R &=& \frac{i}{1+\alpha^2}\left( \gamma \alpha h_z\psi +\alpha4\pi\gamma N_3\frac{1-|\psi|^2}{1+|\psi|^2}\psi\right)\nonumber\\
        &-&\frac{i}{1+\alpha^2}C\frac{1-|\psi|^2}{1+|\psi^2|}\psi
\eeA
and irreversible forces
\beA\label{eq:FI}
F^I&=&\frac{\alpha}{1+\alpha^2}\left( C\frac{1-|\psi_|^2}{1+|\psi^2|}\psi+\gamma h_z\psi\right)\nonumber\\
     &+&\frac{4\pi\gamma N_3}{1+\alpha^2}\frac{1-|\psi|^2}{1+|\psi|^2}\psi,
\eeA
while the baths are the same as in Eq.(\ref{eq:baths}), and the entropy production is given by Eqs.(\ref{eq:phi1}) and (\ref{eq:phi2}).

\subsection{Entropy production in the Frenkel-Kontorova model}%

Let us consider the Frenkel-Kontorova (FK) model, which describes the motion of an oscillator chain sliding over a periodic potential in the presence of random fluctuations:
\beA\label{eq:fkmodel1}
\ddot{x}_m &+& \eta_m\dot{x}_m+g(x_{m+1}+x_{m-1}-2x_m)\nonumber\\
                        &+&h \sin x_m=\sqrt{D_mT_m}\xi_m+b,
\eeA
where for simplicity we consider unit mass oscillators. Here $\eta_m$ the friction parameter, $g$ the coupling strength between the oscillators, $h$ the strength of the on-site potential
and $\xi_m$ a real Gaussian random variable with zero average and variance $\average{\xi_m(t)\xi_m^\prime(t^\prime)}=\delta_{mm^\prime}\delta(t-t^\prime)$.  The diffusion constant reads $D_m=2\eta_m k_B$. 
The last term $b$ is the constant force applied to the chain to make it slide on the periodic potential.

For our purposes, it is useful to introduce the "frequency" $\omega=\sqrt{2g}$ and rewrite the FK equation as

\beA\label{eq:fkmodel2}
\ddot{x}_m &+& \eta_m\dot{x}_m-\omega^2x_m+g(x_{m+1}+x_{m-1})\nonumber\\
                        &+&h \sin x_m=\sqrt{D_mT_m}\xi_m+b
\eeA

Then, one can use the complex coordinates $\psi_m=x_m+\frac{i}{\omega_m}\dot{x}_m$ and get
\beA\label{eq:fkmodel3}
x_m&=&\frac{1}{2}(\psi_m+\psi_m^*)\nonumber\\
\dot{x}_m&=&\frac{\omega_m}{2i}(\psi_m-\psi_m^*).
\eeA

From Eq.(\ref{eq:fkmodel2}) one has that the kinetic term becomes
\be\label{eq:fkmodel4}
\ddot{x}_m=-i\omega\dot{\psi}_m-\frac{\omega^2}{2}(\psi_m-\psi_m^*),
\ee
and finally one obtains
\beA\label{eq:fkmodel5}
\dot{\psi}_m &=&i\omega\psi_m-\eta_m(\psi_m-\psi_m^*)-i\frac{g}{\omega_m}\sin\left(\frac{\psi_m+\psi_nm^*}{2}\right)\nonumber\\
&-&\frac{iA}{\omega}(\psi_{m+1}+\psi_{m-1}+c.c.)\nonumber\\
&+&\frac{iF}{\omega}+\frac{i\sqrt{DT_m}\xi_m}{\omega},     
\eeA
where c.c. indicates the complex conjugate. The complex FK equation  can be obtained as

\be\label{eq:fkhamiltonian1}
\dot{\psi}_m=i\der_m^*\mathcal{H}_{FK}+\frac{i}{\omega_m}\sqrt{DT_m}\xi_m
\ee
where the FK complex Hamiltonian reads
\beA\label{eq:fkhamiltonian2}
\mathcal{H}_{FK} &=& \sum_m \omega\psi_m-i\eta_m\left(|\psi_m|^2-\frac{1}{2}\psi_m^2-\frac{1}{2}\psi_m^{*2}\right)\nonumber\\
		  &+&\frac{2A}{\omega_m}{\Re}\left(\psi_{m+1}+\psi_{m-1}\right)(\psi_m+\psi_m^*)\nonumber\\
		  &+&\frac{b}{\omega}(\psi_m+\psi_m^*).
\eeA
To calculate the entropy production, one needs the irreversible (or dissipative) components of the force, given by $F^I=i\der_m^*\mathcal{H}^I$. It is straightforward to identify the dissipative 
component of the Hamiltonian as $\mathcal{H}^I=-i\eta_m\left(|\psi_m|^2-\frac{1}{2}\psi_m^2-\frac{1}{2}\psi_m^{*2}\right)$. Thus the irreversible force is $F^I=\eta_m(\psi_m-\psi_m^*)$. Then, applying Eqs. (\ref{eq:entropyprod1})
and (\ref{eq:entropyprod2}) gives

\be\label{eq:fkentropy1}
\Phi_{FK} =\sum_m\left(\frac{\average{2\eta_m^2|\psi_n-\psi_m^*|^2}}{D_mT_m/\omega_m^2}-2\eta_m\right).
\ee	    
We remark that, at variance with the DNLS, here the coupling is conservative and does not enter in the definition of entropy production \cite{iubini13,borlenghi17a}.
Finally, we apply the transformations given in Eq.(\ref{eq:fkmodel3}) and go back to the real-valued variables:
\be\label{eq:fkentropy2}	    
\Phi_{FK}=\sum_m\frac{2\eta_m}{k_BT_m}\average{\dot{x}_m^2}-2\eta_m
\ee
The last formula, which contains the particle kinetic energy, is consistent whith what has been obtained in Refs\cite{tome06,tome10} and is the dissipated power. Thus, for standard oscillators, 
using complex or real coordinates gives the same result, as expected. 
Next, we compute the heat flow, defined as the correlation function between reversible and irreversible forces \cite{borlenghi17a}:

\beA\label{eq:heatflux1}
j_{m+1}^Q-j_{m}^Q &=& \frac{2}{\eta_m}{\Re}\average{F_m^IF_m^R}\nonumber\\
			    &=&\average{-i\frac{g}{\omega}(\psi_m-\psi_m^*)(\psi_{m+1}+\psi_{m-1}+c.c.)}\nonumber\\
\eeA
where we identify the heat flow to the correlator between neighbours oscillators

By substituting the expressions for the forces and changing coordinates to the real displacements gives
\be\label{eq:heatflux2}
j_{m+1}^Q-j^Q_{m}=\average{(x_{m+1}+x_{m-1})\dot{x}_m},	
\ee
which is the standard formulation of the heat flow for a chain of oscillators \cite{lepri03,dhar08}.

\section{Conclusions} %

In summary, we have presented a general method, based on stochastic thermodynamics, to calculate entropy production and heat flows in complex-valued Langevin equations with multiplicative noise.
The method is particularly useful to describe the off-equilibrium dynamics of oscillator networks for a variety of physical systems, as described by our examples.
Possible research direction involves formulating the dynamics in terms of a master equation, following the discretisation of the Fokker-Planck equation proposed in Refs.\cite{tome06,tome10}. This should allow
to formulate the irreversibility in terms of fluctuation theorems, relating the synchronisation of the oscillators to the propagating currents and the breaking of detailed balance. These topics will be addressed in future work.

\acknowledgements%

We wish to thank O. Hovorka, A. Silva and S. Iubini for useful discussions. Financial support from Swedish e-science Research Centre (SeRC), Vetenskapsradet (grant numbers VR 2015-04608 and VR 2016-05980), and Swedish Energy Agency (grant number STEM P40147-1) is acknowledged.



\end{document}